\documentclass[aps,pre,twocolumn,groupedaddress]{revtex4-1}
\usepackage{graphicx}
\usepackage{color}

\begin{document}


\title{Anisotropic Particles Strengthen Granular Pillars under Compression}


\author{Matt Harrington}
\affiliation{Department of Physics \& Astronomy, University of Pennsylvania, Philadelphia, PA 19104 USA}
\email{mharrin@sas.upenn.edu}
\author{Douglas J. Durian}
\affiliation{Department of Physics \& Astronomy, University of Pennsylvania, Philadelphia, PA 19104 USA}
\email{djdurian@physics.upenn.edu}



\date{\today}

\begin{abstract}
We probe the effects of particle shape on the global and local behavior of a two-dimensional granular pillar, acting as a proxy for a disordered solid, under uniaxial compression. This geometry allows for direct measurement of global material response, as well as tracking of all individual particle trajectories. In general, drawing connections between local structure and local dynamics can be challenging in amorphous materials due to lower precision of atomic positions, so this study aims to elucidate such connections. We vary local interactions by using three different particle shapes: discrete circular grains (monomers), pairs of grains bonded together (dimers), and groups of three bonded in a triangle (trimers). We find that dimers substantially strengthen the pillar and the degree of this effect is determined by orientational order in the initial condition. In addition, while the three particle shapes form void regions at distinct rates, we find that anisotropies in the local amorphous structure remain robust through the definition of a metric that quantifies packing anisotropy. Finally, we highlight connections between local deformation rates and local structure.
\end{abstract}


\maketitle

\section{Introduction}
\label{sec:Intro}

When a disordered solid is subject to a mechanical load, various characteristics of its local structure and composition directly impact the observed response and performance. For example, composite metallic glasses with interspersed dendrites that arrest shear bands and cracks can counteract the standard trade-off between material strength and fracture toughness in brittle materials~\cite{HofmannNature2008,RitchieNatMater2011}. Other materials can fail in a ductile fashion, in which material failure is marked by local plastic flow and/or growth and coalescence of voids within the bulk~\cite{PINEAU2016424}.

In general, characteristics of the local interactions between constituent elements are critical in determining the response of a disordered system. These descriptors can include bond strength, dissipation, and elasticity. These considerations may require, for instance, additional terms in the development of a constitutive model for the disordered solid, in order to best predict creep and the onset of failure. For example, the Gurson-Tvergaard-Needleman (GTN) model currently serves as a basis for constitutive modeling of ductile failure that can incorporate either void coalescence or plastic flow~\cite{Gurson1977,Tvergaard1981,TvergaardNeedleman1984}. We would like to focus on one aspect that does not inherently alter the interaction between material components, but can still substantially influence behavior: particle shape. 

If the shape of constituent particles (or grains) is changed, that alone may not necessarily alter the inherent physics of how particles interact with one another. The underlying mechanisms of their interactions will remain, but one must consider effects that the shapes have on contact distance, surface curvature, and rotational frustration. Indeed, the effects of grain shapes can be observed in a wide variety of systems, spanning several decades of particle sizes. These phenomena include the toughening of disordered nanoparticle assemblies with elongated particles~\cite{ZhangACSNano2013} and colloidal packings of polygons whose shape frustrates crystalline order~\cite{ZhaoMasonPNAS2015}. On even larger length scales, in which thermal fluctuations are negligible, effects of particle shape become crucially important. Many recent studies have considered the implications of grain shape in granular flows, such as dense driven systems in which nematic ordering can spontaneously occur ~\cite{GalanisSoftMatter2010,BorzsonyiPRL2012,BorzsonyiPRE2012,WegnerSoftMatter2012,BorzsonyiSM2013}, as well as gaseous states in which random collisions impart both translational and rotational motion~\cite{HarthPRL2013}. To better understand the stability of packings of arbitrarily shaped particles, there has also been interest in characterizing (near-)jamming characteristics, such as contact numbers and vibrational modes, of elongated noncircular particles~\cite{PhilipseLangmuir1996,MailmanPRL2009,SchreckSoftMatter2010,ShenPRE2012,SchreckPRE2012,Marschall2017}. 

Recently, grain shapes have been explored as a way to generate free-standing architectural structures~\cite{KellerGM2016}. Examples include highly elongated and U-shaped particles with the capability to form geometrically constrained contacts~\cite{DesmondFranklinPRE2006,StapleGrainsPRL2012,StapleGrainsGM2015} and custom particle fabrication that is facilitated by evolutionary searches for the strongest shapes under a specified load~\cite{PrintingGrainsNatureMat2013,PrintingGrainsSoftMatter2014}. While an overall strength can be prescribed, stress relaxation events, or avalanches, occur as a granular system is slowly driven~\cite{ZapperiPRE1999,MaloneyPRL2004,MaloneyPRE2006,LemaitrePRE2007,MaloneyPRL2009,AmonPRL2012,RegevNatComms2015}. The distribution of sizes of these drops, defined either in terms of a global pressure or energy, often falls on a power law with commonly observed exponents~\cite{DenisovNatComms2016,BaresPRE2017}. Coarse-grained and depinning models have been proposed to associate stress fluctuations with local plastic rearrangements~\cite{DahmenNatPhys2011,Lin07102014}. Particle shape is thought to contribute to the micromechanics of localized slip events~\cite{DahmenNatPhys2011}, but to our knowledge has yet to be explicitly studied within this framework.

When a granular material is slowly driven, it can behave like a slowly deforming solid and provide a bridge to better understanding much of the microscopic behavior within disordered solids. Granular materials are, by definition, assemblies of discrete macroscopic particles, so their constituents can be directly imaged in certain geometries, allowing for a full characterization of microstructure that is not possible in other materials. Other properties of disordered solids, such as bond strength between component particles, can be represented using fluid capillarity~\cite{RieserLangmuir2015} or inter-particle bonding with a cured polymer~\cite{HemmerleSciRep2016}. In this study, we focus on altering particle shapes, with varying amounts of circularity, that comprise a dry granular packing. 

This article is organized as follows. In Section~\ref{sec:methods}, we describe the experimental apparatus, the granular system that is used as a model disordered solid, the particle shapes we study, and the techniques used to collect data on the global and local responses of the material under uniaxial compression. In Section~\ref{sec:Results}, we summarize the primary findings of our study. Specifically, in Section~\ref{subsec:Strength} we describe the effect of particle shapes on the overall material strength, and in Section~\ref{subsec:Avalanches} we discuss stress relaxation events, or avalanches, that occur during compression. Then, in Section~\ref{subsec:Structure}, we describe how we characterize structural anisotropies within the packings through the adaptation of a previously defined metric for non-circular grains. In Section~\ref{subsec:Dynamics} we show how we quantify local plastic strains within the system and test their relationship with avalanches. In Section~\ref{subsec:Connections}, we draw connections between local structure and local dynamics in terms of how anistropies in local plastic strain are correlated with structural anisotropies. Finally, we discuss the broader implications of these findings and motivate further study in Section~\ref{sec:Discussion}.  

\section{Materials \& Methods}
\label{sec:methods}

\begin{figure}
 	\includegraphics[width=0.75\linewidth]{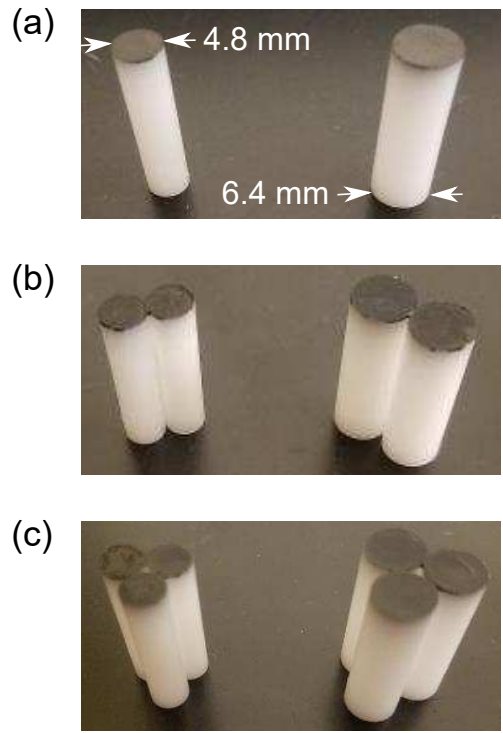}
	\caption{\label{fig:Particles} Photographs of the particle shapes used in this study: (a) monomers, (b) dimers, and (c) trimers. In (a), large and small rods are labeled with their diameters and the height of each rod is 19 mm.} 
\end{figure}

The granular system consists of bidisperse acetal (delrin) rods with diameters 0.25 in (6.4 mm) and 0.1875 in (4.8 mm) and uniform height 0.75 in (19 mm), standing upright on an acrylic substrate. The large and small rods are mixed with a number ratio of 1:1. In order to alter the grain shape, we bond individual rods together to form a composite shape that is overall noncircular, but retains surfaces with a constant radius of curvature. Particle shapes that are comprised of bonded, sometimes overlapping, combinations of circles or spheres is a common technique to explore generic grain shapes, especially in simulation~\cite{ReichhardtEPL2002,SchreckSoftMatter2010,ShenPRE2012,UniaxialNonspherical2012,PrintingGrainsNatureMat2013,PrintingSpheresSoftMatter2014,FranklinShattuck2016,PrintingGrainsSoftMatter2016,Xie2017}, so the technique used here is another iteration of this general approach. While simulations have been used to study the response of circular particles in the apparatus described in this article~\cite{LiRieserPRE2015}, we choose to focus on experiments to establish shape-dependent behaviors before determining how to best incorporate material properties and shape-dependent formulations of contact forces. From here, arbitrary particle number, shape, and size can provide fruitful ventures for simulation study.

\begin{figure}
 	\includegraphics[width=\linewidth]{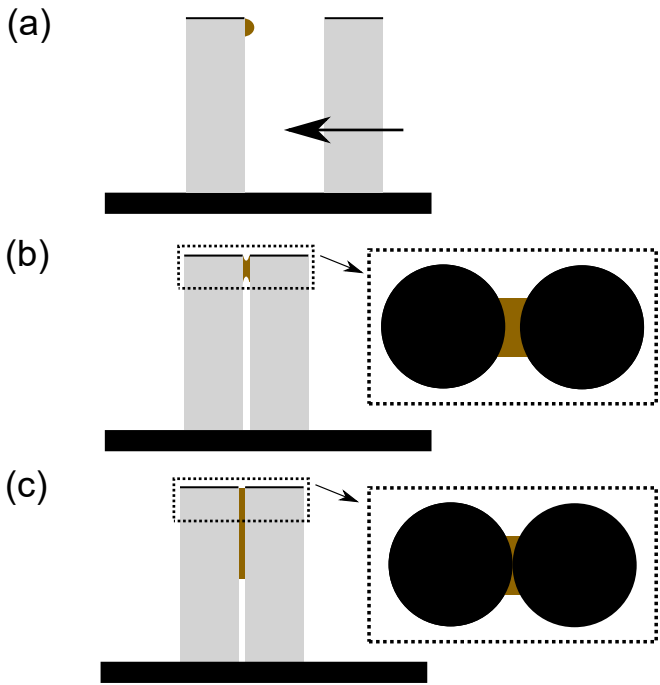}
	\caption{\label{fig:Glue} Schematic of the procedure for constructing dimers. (a) A small drop of adhesive is placed near the top of a rod along its side. A second rod is carefully moved toward the first, both standing on a horizontal table. To ensure both rods are standing upright, they are guided using a vernier caliper set to the rod diameter. (b) The rods are brought nearly into contact, adhesive now adhering to both surfaces. The callout shows a top view. (c) The adhesive spreads between the rods due to capillarity, while simultaneously curing and forming a strong bond. The final separation is exaggerated for clarity. In the top view, the rods are in contact, also shown in Fig.~\ref{fig:Particles}(b). For trimers, a similar procedure is repeated, adding a third rod to form a triangular composite.} 
\end{figure}

The specific shapes we study in this article are monomers (individual plastic rods), dimers (pairs of bonded rods), and trimers (groups of three bonded rods in a triangular shape), as shown in Fig.~\ref{fig:Particles}. Different particle shapes are constructed by gluing rods together using a cyanoacrylate adhesive. The entire fabrication procedure for a dimer is shown in Fig.~\ref{fig:Glue}. A small amount of adhesive is placed near the top of a rod standing upright on a horizontal table. Then, a second rod, also standing upright, is brought into contact with the first. To ensure both rods are straight, they are confined to stand within the jaws of a vernier caliper set to the rod diameter. The adhesive spreads down the pair of rods through capillarity, while also curing to form a strong bond between the rods. The amount of adhesive used is not precise, but it must be substantial enough so that the cured bond is strong, yet limited so the adhesive does not spread all way down the rods, bonding them to the table. When fully cured, the pair of rods now form a dimer. To make a trimer, this same adhesive procedure is repeated with a third rod brought in to form a triangle. This type of trimer is preferred, as a linear chain of more than two tall macroscopic rods is generally difficult to achieve by hand with sufficient accuracy and consistency. Furthermore, this allows us to isolate dimers as our case study in elongated particles, while the trimers are more axially symmetric, but with characteristic bumps. After allowing the adhesive to cure overnight, the dimers and trimers require substantial effort to break apart by hand.  Without precisely measuring shear and/or flexural strength, we observe that internal stresses within each experiment never cause breakage.

\begin{figure}
 	\includegraphics[width=0.75\linewidth]{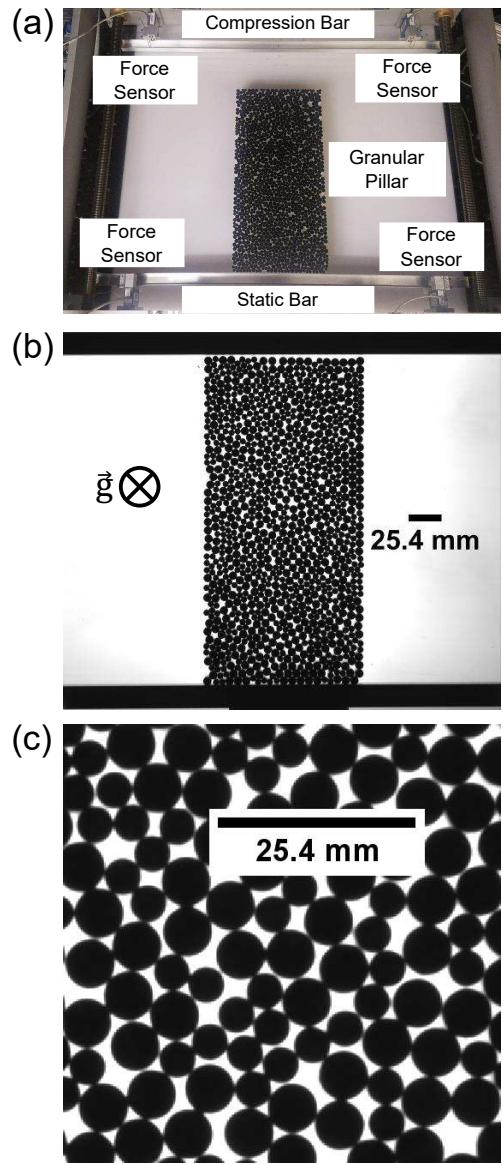}
	\caption{\label{fig:App} (a) A photograph of the apparatus from the top-down, slightly off-line from the actual camera used to image the system. The acquisition system and the stepper motor that drives the compression bar are not pictured. (b) A raw unprocessed image of the granular pillar comprised of dimers. The compression and static bars are visible in the top and bottom of the image, respectively. The direction of gravity is also labeled. (c) A close-up image of a neighborhood of dimers within the interior of the pillar.}
\end{figure}

Our experimental apparatus is shown in Fig.~\ref{fig:App}(a), with its various components labeled. This is the same apparatus used in Refs.~\cite{CubukSchoenholzPRL2015,RieserLangmuir2015,LiRieserPRE2015,RieserPRL2016,RieserThesis} to study granular pillar deformation. The entire apparatus lies on a horizontal table-top, so gravity does not directly drive or hinder the motion of grains. The grains, all of one chosen shape, are arranged into a tall, narrow pillar with an aspect ratio of approximately 2:1 using a rigid frame. An initial pillar configuration is shown in Fig.~\ref{fig:App}(a)-(b). Since the particle shapes have distinct area fractions when packed randomly, we choose to keep the width of the pillar consistent ($W_0$ = 4.875 in = 12.4 cm), while the initial pillar height ($H_0 \sim$ 9.75 in = 24.8 cm) can vary slightly from one trial to another, much less from one shape to another. Differences in $H_0$ between trials are especially apparent in pillars comprised of dimers and trimers, since the particle geometries frustrate random close-packing as opposed to the circularly symmetric monomers. The initial area fraction of monomers is $\phi = 0.823 \pm 0.004$, dimers, $\phi = 0.809 \pm 0.007$, and trimers, $\phi = 0.805 \pm 0.005$. The uncertainties in $\phi$ are determined from the range covered over all trials.

The pillar is unaxially compressed from the top by a slowly moving bar ($v_c$ = 0.033 in/s = 85 $\mu$m/s), while a static bar remains in contact with the pillar bottom. As the pillar is compressed and laterally spreads out, its interior structure constantly evolves due to interspersed local plastic flow and the creation and collapse of voids. These aspects are commonly present in materials undergoing ductile failure~\cite{PINEAU2016424}, so our apparatus can serve as a model system for this type of material failure. An important distinction between this apparatus and other uniaxially driven granular systems~\cite{UniaxialNonspherical2012,UniaxialNonspherical2016} is that we do not restrict expansion of the system with any sort of hard boundary or soft membrane, nor is the compression direction along or against the direction of gravity. We performed 5 trials for each type of pillar composition, with the specifics of the microscopic initial structure varying from run to run, but initial dimensions remaining constant as described above. Keeping the system dimensions consistent across shapes also requires altering the total number of particles. $N = 1000$ for monomers, $N = 500$ for dimers, and $N = 334$ for trimers. We chose to keep the pillar size constant, rather than the discrete particle count, in order to draw fair comparisons of material strength and behavior. In fact, large pillars comprised of $1000$ dimers or $1000$ trimers would present practical challenges for the present apparatus. Similar studies that can control for pillar size, particle count, and particle mass, via simulation or custom particle fabrication, would make for interesting studies.

While the compressing bar is in motion, we acquire 4.2 megapixel (2048x2048) images of the pillar deformation using a JAI/Pulnix TM-4200CL camera with a frame rate of 8 fps. For each image, we simultaneously record the forces exerted on the moving and static bars using Omega Engineering LCEB-5 force sensors. After acquiring images, we locate all circular particles using a circular edge-finding algorithm~\cite{RieserThesis}. Fig.~\ref{fig:App}(c) demonstrates the sharp intensity contrast between the painted caps of the particles and the background illumination. The displacement of the compressing bar between successive frames is about $10^{-3}R$, where $R$ is the large monomer radius, so linking position coordinates together into particle tracks is a straightforward process. To suppress noise in particle positions, we apply a Gaussian filter with a time window equal to the time over which the compression bar moves $\frac{2}{15}R$. This becomes the effective time interval between filtered frames. We also use this Gaussian smoothing to differentiate positions, yielding approximations of instantaneous velocities. 

When analyzing pillars with dimers or trimers, we group rods together by measuring interparticle distances over time. Since every dimer and trimer consists of equal sized rods, we can deduce some of the combinations just from the initial packing. Within portions of the pillar that significantly deform over the full run (over which the bar moves about halfway down the initial height), we usually find there is only one possible combination to link dimers or trimers together. For regions that do not significantly deform, especially large clusters of like-sized particles at the bottom of the pillar, we group particles such that interparticle distance fluctuations are minimized. Ultimately, we can successfully group every dimer and trimer together, particularly those that exhibit motion beyond our noise level in calculating positions. The centroid positions of the dimers and trimers are directly calculated from the positions of their constituent rods, smoothed, and differentiated as described above.

After particle tracking and the identification of monomers, dimers, and trimers, we can measure various aspects of local structure and motion. These shall be discussed in further detail in Sections~\ref{subsec:Structure} and~\ref{subsec:Dynamics}.

\section{Experimental Results}
\label{sec:Results}

\subsection{Material Strength}
\label{subsec:Strength}

\begin{figure}
 	\includegraphics[width=\linewidth]{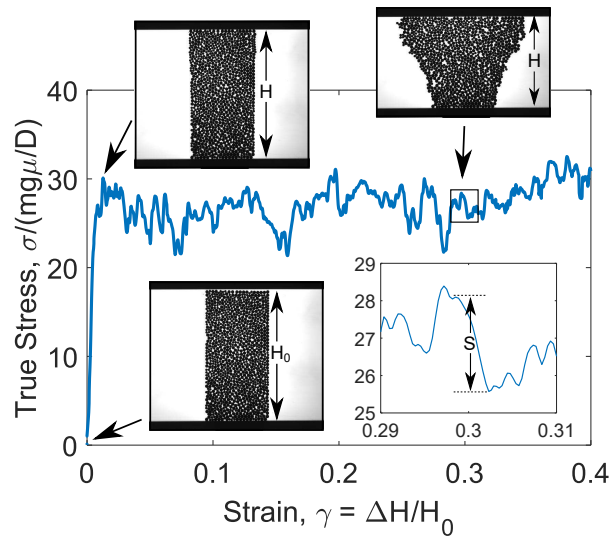}
	\caption{\label{fig:sampleStressStrain} (Main) Stress-strain curve for a single trial of compression of a pillar with dimers. The stress is the force applied by the moving bar divided by the current width of the pillar, $\sigma = F/W$. The strain is the change of the pillar height divided by the initial pillar height, $\gamma = \Delta H/H_0 = (H_0-H)/H_0$. Raw images of the pillar are interspersed along the curve, corresponding to points when the pillar is in its initial condition ($\gamma = 0$), at yield ($\gamma \sim 0.01$), and undergoing long-term deformation and failure ($\gamma > 0.01$). (Inset) A zoomed-in area of the failure portion of the stress-strain curve (outlined with a black box in the main plot) exhibiting several avalanches. The largest avalanche in this window is labeled $S$.}
\end{figure}

As an analog to standard tests of material strength, we measure the stress-strain response of the pillar as it is compressed. Following the procedure set in Ref.~\cite{LiRieserPRE2015}, we quantify the compressive stress, $\sigma$, as ${F}/{W}$ where $F$ is the driving force exerted by the moving bar on the pillar and $W$ is the current width of the pillar in contact with the moving bar, making $\sigma$ a measurement of true stress. A rod is considered to be in contact with the moving bar if its vertical position is within $0.25R$ of the rod at the top of the pillar, where $R$ is the large rod radius. The pillar width $W$ is calculated as the end-to-end horizontal distance of these contacting rods. The forces on the static bar are negligible for monomer runs, so to be consistent across all trials we choose to focus on just the force actively driving the pillar. As expected, both $F$ and $W$ tend to continuously increase over the course of a pillar compression. While $W$ tends to grow steadily over the course of a compression, it can exhibit a jump discontinuity if new particles(s) come into or out of contact at either end of the pillar top, while new contacting particles within the interior of the pillar top, the primary mechanism of width increase, do not result in $W$ discontinuities or fluctuations. Indeed, jumps in $W$ only occur about 5 times in a single run, so it primarily behaves as a smooth function without inducing substantial fluctuations in stress, $\sigma$. In every plot showing stress, we quantify $\sigma$ in derived units of $mg\mu/D$, where $m$ and $D$ are the mass and diameter, respectively, of a large rod, $g$ is the acceleration due to gravity, and $\mu$ is the grain-substrate coefficient of friction, measured to be $0.23 \pm 0.01$~\cite{RieserThesis}. Effectively, these units represent the stress required to move an individual large monomer at constant speed. The vertical strain $\gamma$ is given by ${\Delta H}/{H_0}$, where $H_0$ is the initial height of the pillar and $\Delta H$ is the difference in height between the initial pillar and the deformed system, $H_0 - H$. 

In Fig.~\ref{fig:sampleStressStrain}, we show the stress-strain behavior for a single compression trial of a pillar comprised of dimers and highlight three regimes of pillar deformation: (1) an elastic-like initial compression, which occurs over a very short strain range ($\gamma \lesssim 0.01$), too short to confirm a linear response, (2) a yield transition around $\gamma \sim 0.01$ when stress reaches a maximum value, and (3) long-term ($\gamma \gtrsim 0.01$) deformation and failure that is marked by a fairly constant material strength, with irregular stress fluctuations. In Section~\ref{subsec:Avalanches} we will consider the distribution of stress drops, but for now we are motivated by their relative size and irregular frequency to consider trial averages as a way of better gauging the material strength of pillars comprised of our three particle shapes. 

\begin{figure}
 	\includegraphics[width=\linewidth]{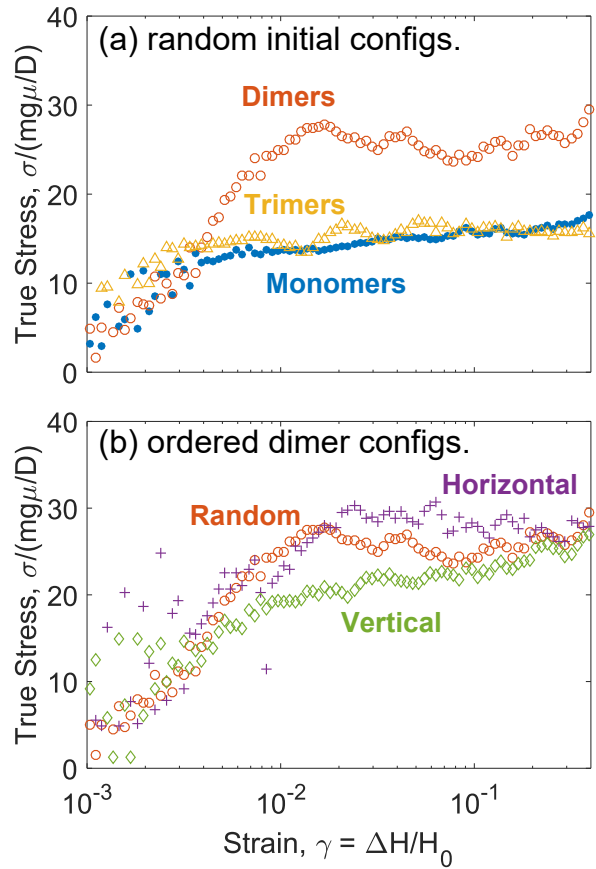}
	\caption{\label{fig:Strength} Stress-strain curves for (a) pillars comprised of randomly packed shapes: monomers (closed circles), dimers (open circles), and trimers (triangles), and (b) different preparation protocols of dimers: random (open circles), horizontal ($+$'s), and vertical (diamonds). The results shown in these plots are derived from 5 trials averaged together. Strain is shown on a log scale, in order to highlight both the  modulus at strains belong yield and the long-term material strengths for strains well beyond yield. Before averaging the point of zero strain $\gamma = 0$ is adjusted to minimize contributions from individual particle motions at the top of the pillar.}
\end{figure}

We average 5 trials together to generate stress-strain curves, shown in Fig.~\ref{fig:Strength}, significantly reducing the prevalence of stress fluctuations during long-term deformation. Note that Fig.~\ref{fig:Strength} is presented with a horizontal log scale, emphasizing low-strain behavior. Before averaging, the point of zero strain, $\gamma = 0$, in each trial is set to minimize initial strain readings that result from the motion of individual particles within the top layer of particles. 

We see in Fig.~\ref{fig:Strength}(a) that dimers exhibit more strength than monomers, in terms of a compressive modulus that can be estimated from the quasi-elastic regime, a larger yield stress, as well as the stress required to continually deform the pillar at large strains. Pillars comprised of trimers retain an average long-term strength that is comparable to that of monomers.  

\begin{figure}
 	\includegraphics[width=\linewidth]{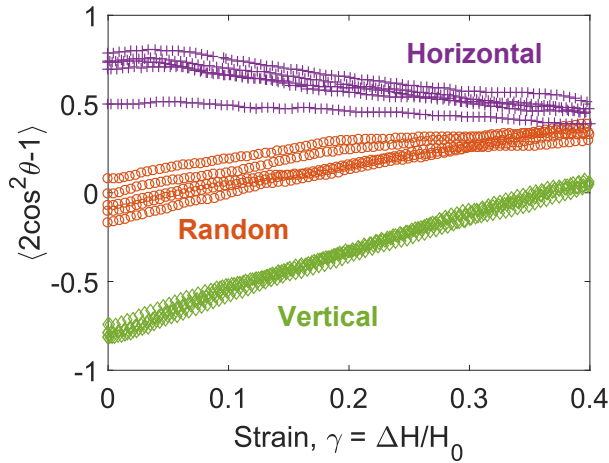}
	\caption{\label{fig:OrderParam} Evolution of a two-dimensional order parameter, $\langle 2\cos^2 \theta - 1\rangle$, for every dimer packing. $\theta$ is the angle between the orientation of each dimer and the horizontal compressing bar and $\langle \cdot \rangle$ is the ensemble average. The order parameter value at $\gamma = 0$ indicates the amount of order present in the initial state, either along or perpendicular to the compression bar.}
\end{figure}

Dimers are clearly the strongest shape tested in this study, so we would like to further investigate why this is the case. The specific question we would like to answer is: can we prepare a pillar using dimers in a way that either strengthens the pillar to a further degree or diminishes the apparent strengthening effect? To do so, we note the unidirectional driving of the system, in junction with the elongation of the dimers, to prepare two types of highly ordered packings of dimers. In addition to the disordered dimers previously measured, we prepare a set of packings in which dimers are preferentially ordered horizontally, along the compressing and static bars, as well as a set of packings with dimers preferentially ordered vertically, along the compression direction. These pillars are meticulously created layer-by-layer, building upward in the horizontal case and to the right in the vertical case, in an effort to minimize the presence of orientational defects. The pillar dimensions are kept consistent as before, which necessitates the presence of some defects. Due to the high degree of ordering, the initial packing fractions for ordered dimers is higher in both cases, with $\phi = 0.813 \pm 0.002$ for horizontal dimers and $\phi = 0.814 \pm 0.003$ for vertical dimers. We also quantify the degree of orientational order present in the initial pillars, and during compression, as shown in Fig.~\ref{fig:OrderParam}.

\begin{figure*}
 	\includegraphics[width=\textwidth]{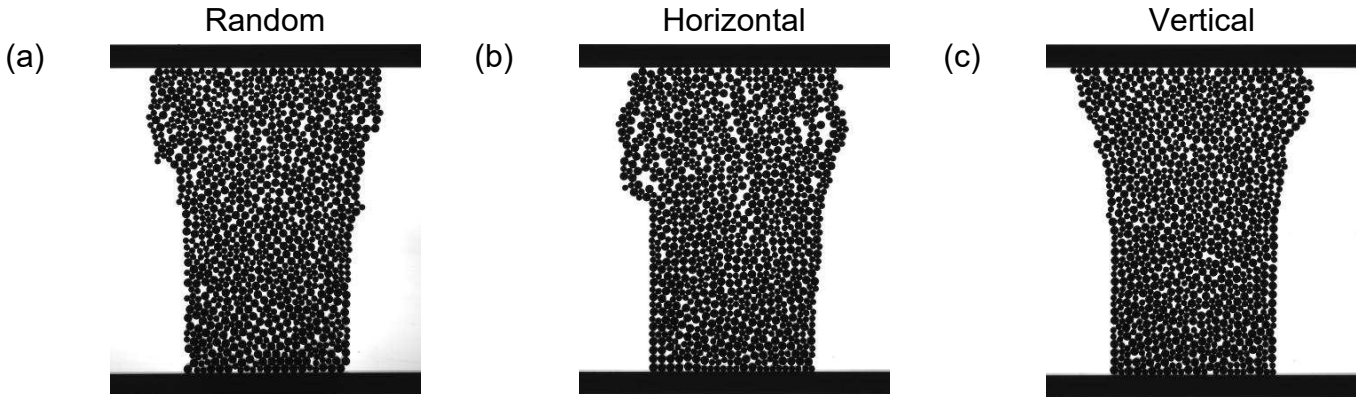}
	\caption{\label{fig:Ordered_Dimers} Snapshots of pillars comprised of dimers undergoing deformation at the same strain ($\gamma \sim 0.1$). Each picture corresponds to a different orientational packing protocol: (a) random, (b), horizontal, and (c) vertical. Note the buckling behavior in (b), as well as the smooth pillar boundaries in (c). Full movies including overlays with $D^2_{min}$ can be found in the Supplementary Material~\cite{Supplement}.}
\end{figure*}

We note a marked distinction in the material response for these three types of dimer packings, illustrated in Fig.~\ref{fig:Strength}(b). Specifically, vertical dimers are substantially weaker than the randomly packed dimers. From there, we see that horizontal dimers reach an even higher compressive strength at $\gamma \sim 0.02$.  The presence of noise in the low-strain behavior of the ordered pillars should be noted. This noise can be attributed to the presence of orientational defects, specifically those near the top of the pillar, in individual trials. These compound the difficulty of differentiating low strain behavior to define a compressive modulus. Nevertheless, the pillar strength is substantially impacted not only by the grain shape, but also the procedure by which the packing is generated. 

Looking at raw snapshots of these packing types under compression, illustrated in Fig.~\ref{fig:Ordered_Dimers}, we also observe distinct local behaviors as the dimers are compressed. These differences are also apparent in full movies in the Supplementary Material~\cite{Supplement}. The movies in the Supplementary Material include overlays with $D^2_{min}$, a metric that quantifies plastic deformation and local rearrangement around each discrete particle~\cite{FalkLangerPRE1998}. We overlay with $D^2_{min}$, which is assigned to individual particles, rather than $J_2$, a measurement of local instantaneous strain rate (discussed at length in Section~\ref{subsec:Dynamics}), which is defined for regions of three particles. $J_2$ overlays would thus obscure dimer positions and orientations. The horizontal dimers buckle outward, breaking into separate columns with little slip between particles, as shown in Fig.~\ref{fig:Ordered_Dimers}(b). Also, the shape of the pillar expands with rough edges, the furthest outward extents lying about a quarter of the way down the pillar. Meanwhile, the vertical dimers deform much more gradually, shown in Fig.~\ref{fig:Ordered_Dimers}(c) with a smooth symmetric plume right at the very top of the pillar. When dimers are packed randomly, as in Fig.~\ref{fig:Ordered_Dimers}(a), contributions from both types of deformation are present. The amount of structural rupture occurring within the interior of the pillar is quantified in Section~\ref{subsec:Structure}.

We can now state that the material strength gained from dimer packings comes directly from dimers that preferentially lie ordered to each other, specifically interlocking along the horizontal direction as to resist outward expansion of the pillar. We can even see in the right side of Fig.~\ref{fig:Strength}(b), in junction with Fig.~\ref{fig:OrderParam}, that as the random and vertical dimer pillars are continually deformed, dimers rearrange so that those in contact with the bar are mostly horizontal, while the strength of the pillar continually increases. In fact, they are trending toward the strength exhibited by pillars with horizontal dimers to begin with. This result provides further motivation to investigate the relationship between local structural and deformation features, which shall be discussed in Sections~\ref{subsec:Structure},~\ref{subsec:Dynamics}, and~\ref{subsec:Connections}.

\subsection{Avalanches \& Stress Relaxation}
\label{subsec:Avalanches}

\begin{figure}
 	\includegraphics[width=\linewidth]{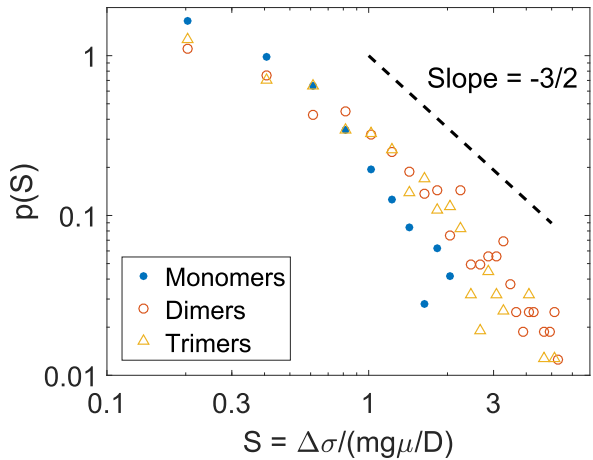}
	\caption{\label{fig:AvalancheDists} Distributions of avalanche sizes $S = \Delta\sigma/(mg\mu/D)$ across all trials for monomers, dimers, and trimers. An increasing portion of the distributions, for values below the noise level of stress, is not shown. It is difficult to assign fits given the lack of range, though it seems all curves would fit a power law with exponent $-3/2$ reasonably well. We also see monomers have a smaller maximum $S$ compared to dimers and trimers.}
\end{figure}

As previously mentioned in Section~\ref{subsec:Strength}, the stress-strain curve for each compression trial exhibits fluctuations about an average strength during the regime of large strain. The same trend is seen in other amorphous systems, the mechanism of which can be owed to the build up of local stresses, followed by a relaxation that is associated with slip rearrangements~\cite{ZapperiPRE1999,DahmenNatPhys2011,Lin07102014,RegevNatComms2015,DenisovNatComms2016}. In this section, we consider the sizes of stress relaxation events, or avalanches, and their frequency as a function of particle shape. Later, in Section~\ref{subsec:Dynamics}, we will consider potential origins of the stress fluctuations at a more local scale.  

To be clear, when we refer to avalanches and their sizes, we are exclusively referring to continuous drops in stress, as illustrated in Fig.~\ref{fig:sampleStressStrain}. Avalanches are generally preceded by a build-up of stress within the system, which the avalanche at least partially relaxes away. It is worth noting that the representative data in Fig.~\ref{fig:sampleStressStrain} includes stress accumulations and avalanches that are roughly symmetric with respect to strain. This aspect of symmetry is likely due to the hardness of rods preventing elastic energy from being stored locally, along with the lack of a confining boundary permitting the pillar to constantly dilate globally. In each individual run, we locate intervals over which the stress is decreasing, truncated by peaks, valleys, and/or plateaus in $\sigma$. The dimensionless magnitude of the stress difference of the entire interval is defined to be $S = \Delta\sigma/(mg\mu/D)$. In Fig.~\ref{fig:AvalancheDists}, we show the distributions of $S$ for the three particle shapes, measured over all trials. The lower bound of the plotted range in $S$ is chosen to neglect a region where the distributions are increasing, which coincides with avalanches that are below the noise level in our stress measurements. 

As we expect from other studies of avalanche distributions within amorphous systems, we see that the distributions could be described by a power law. In fact, the exponent for distributions over the range $S>1$ is approximately $-3/2$, which has been observed in other amorphous systems~\cite{KablaJFM2007,DenisovNatComms2016} and predicted by a coarse-grained model~\cite{DahmenNatPhys2011}. We should note that, while we are estimating $-3/2$ as the exponent, we cannot confidently calculate this exponent given the narrow range of $S$. This is due to both the noise level in measuring stress, as well as substrate friction, as the maximum observed value for all avalanche sizes is determined by the force required to move $\mathcal O(10)$ particles. Furthermore, while the applicability of power laws in other amorphous systems motivates the conjecture of a $-3/2$ power law, we find that the complementary cumulative distribution function of avalanche sizes can be fit over its full range with a compressed exponential function, $\exp(-(S/S_0)^\beta)$. 

While the avalanche distribution for all shapes have approximately the same rate of decay, Fig.~\ref{fig:AvalancheDists} shows unique features of the distributions. Monomers exhibit smaller avalanches, while the distributions for dimers and trimers are similar. This is also reflected in compressed exponential fits for the complementary cumulative distribution function. Monomers have $\beta = 1.5 \pm 0.2$ and $S_0 = 0.89 \pm 0.05$, dimers have $\beta = 1.4 \pm 0.2$ and $S_0 = 2.4 \pm 0.2$, and trimers have $\beta = 1.4 \pm 0.2$ and $S_0 = 1.7 \pm 0.1$. In Ref.~\cite{BaresPRE2017}, increased particle friction is observed to result in larger upper thresholds in avalanche size. Since the bumpy concave shapes of dimers and trimers effectively increase particle friction, we observe a similar trend. We do not show the avalanche distributions for the highly ordered dimer packings, as they are virtually identical to the avalanche distribution of randomly packed dimers.

Particle shape thus directly influences the global material response, both in terms of averages and fluctuations of stress. In Sections~\ref{subsec:Structure},~\ref{subsec:Dynamics}, and~\ref{subsec:Connections}, we further explore the effects of particle shape on both local structure and dynamics.

\subsection{Local Structure}
\label{subsec:Structure}

While the granular pillars are initially set with consistent dimensions, there are bound to be heterogeneities in the packing efficiency, much less additional structural heterogeneities that are introduced as the pillar is compressed. Furthermore, voids that form or collapse over time are crucial componments of ductile failure. To quantify these aspects of local structure, we use the dimensionless quantity $Q_k$, previously defined in Ref.~\cite{RieserPRL2016}, to highlight anisotropies in the Voronoi tessellation of the packing.  In the simple case of monomers, we perform a radical Voronoi tessellation of the particle positions using the software package {\tt voro++}~\cite{RycroftChaos2009}, followed by a Delaunay triangulation. Then, we define a vector field $\mathbf{C}$ that points from the rod center to the centroid of its own Voronoi cell. Finally, we define $Q_k$ for a triangle $k$ from the divergence of this vector field,
\begin{equation}
\label{eq:Qk}
	Q_k = \nabla \cdot \mathbf{C}_k \frac{A_k}{\langle A \rangle},
\end{equation}
where $A_k$ is the area of triangle $k$ and $\langle A \rangle$ is the average area of all triangles. Scaling the divergence by area sets $\langle Q_k \rangle = 0$, with some residual contribution from the finite boundaries of the experimental data. To minimize these boundary effects, we ignore all triangles that lie on the boundary of the pillar. $Q_k$ is highly correlated with relative free area fraction, where $Q_k < 0$ corresponds to under-packed regions, while $Q_k > 0$ corresponds to over-packed regions. Furthermore, the distribution of $Q_k$ values measured for either experimental hard disks or simulated soft disks is nearly Gaussian and centered at $Q_k = \langle Q_k \rangle = 0$, in sharp contrast to distributions of local free volume. The deviation from Gaussianity in the tail of $Q_k$ indicates a surplus of underpacked particles, with both standard deviation and skewness of $Q_k$ exhibiting kinks at the jamming point $\phi_c$~\cite{RieserPRL2016}. 

\begin{figure}
 	\includegraphics[width=\linewidth]{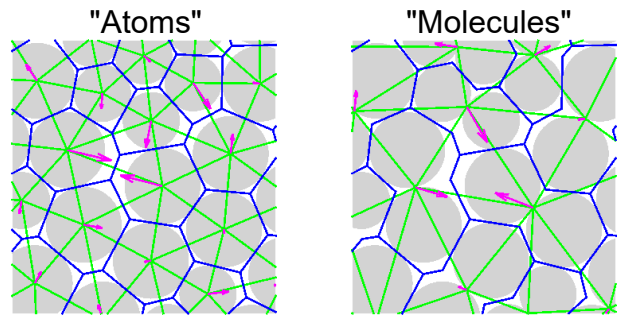}
	\caption{\label{fig:Voronoi} Illustration of two approaches for probing the local structure of the pillar, using data from a packing of dimers. Voronoi tesselation and Delaunay triangulation are drawn with respect to either (left) individual rod positions (``Atoms") or (right) composite particle centroids (``Molecules"). The ``molecular" approach takes the same Voronoi tessellation produced using the ``atomistic" approach and cuts out edges drawn across dimer and trimer bonds. In both, the arrows $\mathbf{C}$ point from the center of the rod/particle to the centroid of its Voronoi cell (magnified $10\times$).}
\end{figure}

\begin{figure}
 	\includegraphics[width=\linewidth]{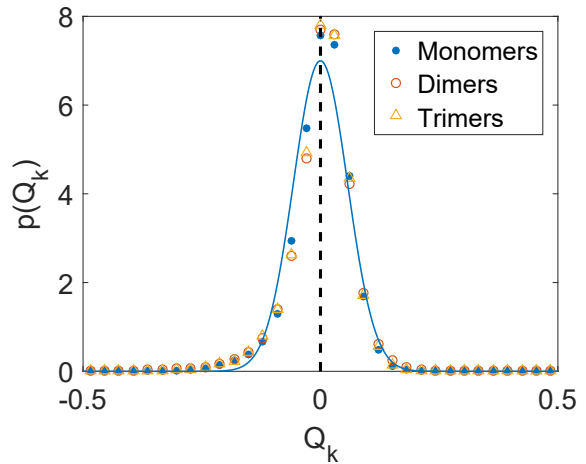}
	\caption{\label{fig:Qk_ALL_Lin} Linear plot of the probability density functions of $Q_k$, defined as the weighted divergence of the particle center-to-cell centroid vector field in Equation~\ref{eq:Qk}, derived from the ``molecular'' approach, for deformed random packings of the different particle shapes. A triangle over which $Q_k$ is calculated is considered deformed if at least one of its constituent particles has moved at least one large monomer radius $R$. A dashed line is drawn at $Q_k = 0$, separating under-packed ($Q_k < 0$) and over-packed ($Q_k > 0$). The solid line is an ideal Gaussian curve for monomers, given the mean and standard deviation of $Q_k$. The semilog version of this plot is shown in Fig.~\ref{fig:Qk_ALL}(b).}
\end{figure}

Calculating $Q_k$ with the centroid positions of dimers and trimers requires a small amount of adaptation in the method, as performing the Voronoi tessellation of non-spherical particles can often result in non-convex Voronoi cells~\cite{SchallerEPL2015}. Fortunately, given that the dimers and trimers both have circular curvature, we can rely heavily on the initial Voronoi analysis. Starting with the Voronoi tessellation for rods generated from {\tt voro++}, we can simply delete edges that cut across bonded particles. This leaves a larger effective cell that now surrounds the entire dimer pair or trimer group. A new triangular tessellation is then computed, using knowledge of particles that share Voronoi edges. Fig.~\ref{fig:Voronoi} illustrates the two approaches that can be used for computing $Q_k$ for a region of dimers. While the triangulation of dimers and trimers is no longer dual with its Voronoi diagram, this remains an effective way to determine a packing tessellation with no gaps or overlaps. Moreover, Fig.~\ref{fig:Qk_ALL_Lin} indicates that $Q_k$ measured in this ``molecular'' sense retains a Gaussian-like profile on a linear scale. When characterizing local structure in the dimer and trimer packings, we have actually found both pictures can be enlightening: one where $Q_k$ is calculated based on individual rod positions (``Atoms") and one where we instead use the centroid of the composite shape (``Molecules").

\begin{figure*}
 	\includegraphics[width=0.75\textwidth]{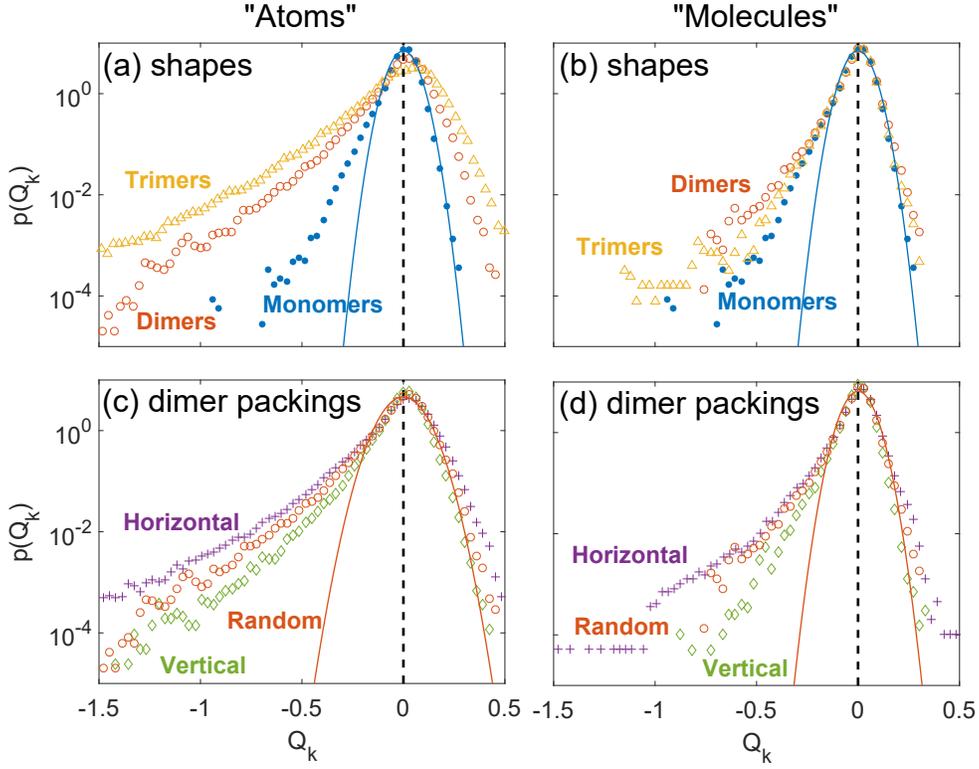}
	\caption{\label{fig:Qk_ALL} Semilog plots of the probability density functions of $Q_k$, defined as the weighted divergence of the particle center-to-cell centroid vector field in Equation~\ref{eq:Qk}, for (a),(b) deformed random packings of the different particle shapes and (c),(d) deformed dimer packings with different initial ordering protocols. The ``atomistic'' approach is used in (a),(c), while the ``molecular'' approach is used in (b),(d). A triangle over which $Q_k$ is calculated is considered deformed if at least one of its constituent particles has moved at least one large monomer radius $R$. In every plot, a dashed line is drawn at $Q_k = 0$, separating under-packed ($Q_k < 0$) and over-packed ($Q_k > 0$). Solid lines are ideal Gaussian curves for monomers in (a),(b) and random dimers in (c),(d), given the mean and standard deviation of $Q_k$.}
\end{figure*}

The ``atomistic" $Q_k$, illustrated on the left side of Fig.~\ref{fig:Voronoi}, highlights absolute areas of vacancies and has been shown to correlate well with local free area~\cite{RieserPRL2016}. Fig.~\ref{fig:Qk_ALL}(a) shows the probability density functions for $Q_k$ measured in regions that have been driven at least one large monomer radius from its initial position. From monomers to dimers to trimers, we see that using larger shapes results in distinctly larger voids during deformation. At the same time, bonded rods also allow for additional regions that are overpacked, especially when a triangle corresponds to a discrete trimer particle. These effects are plainly visible by eye in the raw experimental data and are quantified using this method of $Q_k$ measurement.

However, relative structural anisotropies are less apparent when accounting for both the shape and orientation of the discrete particles. While the dimers and trimers create large voids, are they consistent with the fact that dimers and trimers are themselves larger? Another question lies in whether similarities in the random preparation protocol for all shapes can be captured in a structural quantity. These questions can be addressed by measuring the ``molecular" $Q_k$, illustrated on the right side of Fig.~\ref{fig:Voronoi}, with distributions shown in Fig.~\ref{fig:Qk_ALL}(b). Remarkably, despite the randomness of dimer and trimer packings resulting in more physical void space, we see that the $Q_k$ distributions are strikingly similar. All the distributions are nearly Gaussian in the vicinity of $Q_k = \langle Q_k \rangle = 0$ and retain similar widths despite the manifestation of distinct global dilation rates. The collapse of these distributions suggests that $Q_k$, as a metric for local packing anisotropy, may serve well beyond characterizing local free area in packings of circles. Rather, $Q_k$ seems to demonstrate promise to characterize local packing structure with arbitrary particle shape, and that random close packings of symmetric and asymmetric particles can exhibit similar local structural fluctuations.

To quantify the collapse of $Q_k$ distributions that results from moving from the ``atomistic'' picture to the ``molecular'' picture, we compute the skewness and kurtosis of $Q_k$ distributions shown in Table~\ref{table:QkMoments}. Indeed, similar values are reported for monomers and the ``molecular'' dimers and trimers. It is also worth noting the physical interpretations of skewness and kurtosis in the context of $Q_k$. Skewness provides a measurement of the asymmetry of a distribution, while kurtosis quantifies the presence of tails, either fat or broad relative to a Gaussian distribution. While $Q_k$ appears near Gaussian in the linear plots shown in Fig.~\ref{fig:Qk_ALL_Lin}, there are necessary deviations in its skewness and kurtosis. For one, there is a finite limit to how closely hard particles, such as the ones used in this study, can pack together, while void space in underpacked regions is only restricted by the boundaries of the system, which in this case are open. This allows a wider accessible range in negative $Q_k$ values, resulting in a negative skewness. Figs.~\ref{fig:Qk_ALL_Lin} and~\ref{fig:Qk_ALL}(b) illustrate this asymmetry, since the empirical data in the left tail for monomers lies slightly above the ideal Gaussian curve, while the right tail more closely follows the ideal curve. In turn, the wider range of negative $Q_k$ values requires its tail to decay slower than the Gaussian curve, which is apparent throughout Fig.~\ref{fig:Qk_ALL}. Hence, the kurtosis of $Q_k$ will be higher than that of a Gaussian. As expected, these aspects of $Q_k$ are reflected in Table~\ref{table:QkMoments} for the collapsed ``molecular'' distributions.

\begin{table}[h]
\centering
\begin{ruledtabular}
\caption{Skewness and kurtosis of $Q_k$ distributions shown in Fig.~\ref{fig:Qk_ALL}(a),(b). All calculations are restricted to the range $-0.25 < Q_k < 0.25$, to highlight fluctuations in $Q_k$ near $\langle Q_k \rangle = 0$ and reduce the impact of low frequency outliers. For reference, the skewness of a Gaussian distribution is $0$ and its kurtosis is $3$.}
\begin{tabular}{ccc}
 Shape &  Skewness &  Kurtosis \\ \hline
 Monomers & $-0.68$ &  $4.4$\\ 
 Dimers (``Atoms'') & $-0.36$ &  $3.3$\\ 
 Trimers (``Atoms'') & $-0.26$ &  $2.5$\\ 
 Dimers (``Molecules'') & $-0.63$ &  $4.6$\\ 
 Trimers (``Molecules'') & $-0.71$ &  $4.5$\\ 
\end{tabular}
\label{table:QkMoments}
\end{ruledtabular}
\end{table}


For a particular non-circular shape, $Q_k$ can also indicate distinct structural characteristics. Fig.~\ref{fig:Qk_ALL}(c)-(d) shows $Q_k$ distributions for the different dimer packings, using the ``atomistic'' and ``molecular'' views of $Q_k$ for deformed regions. As previously suggested in Section~\ref{subsec:Strength}, horizontally ordered dimers strengthen the pillar, while also giving way to additional local rupture. While ordered dimers are initially packed with similar global area fractions, Fig.~\ref{fig:Qk_ALL}(c)-(d) indicates the formation of additional void space when dimers are initially packed horizontally. Vertically packed dimers form voids at a more gradual rate, while randomly packed dimers lie at a rate between the two ordering procedures.

While the $Q_k$ distributions for different particle shapes collapse very well in Fig~\ref{fig:Qk_ALL}(b), it is worth nothing that some deviation is seen for highly under-packed regions, where $Q_k \lesssim -0.3$. This kink is even exacerbated in the case of horizontally ordered dimers, shown in Fig.~\ref{fig:Qk_ALL}(d). To seek a dynamical explanation for this feature, we now shift our attention to local deformation.

\subsection{Local Dynamics}
\label{subsec:Dynamics}

In addition to local structure, we can also quantify local plastic strain within the pillar, another important feature of ductile failure. In this study, we choose to quantify local deformation by the deviatoric strain rate, $J_2$, which describes how the shape of a small region deforms. The procedure of calculating $J_2$ is as follows.

\begin{figure}
 	\includegraphics[width=\linewidth]{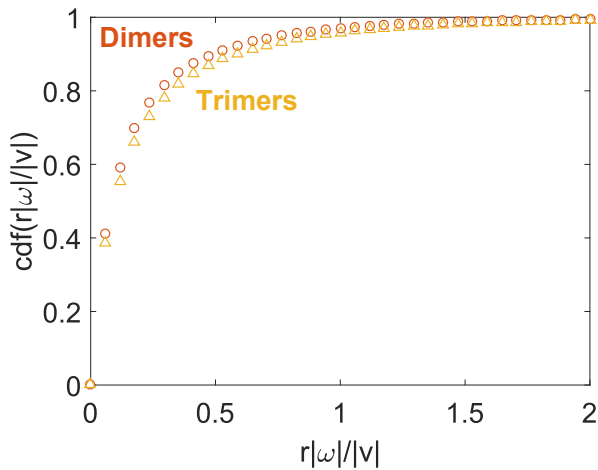}
	\caption{\label{fig:cdf_rotFrac} Cumulative distribution functions of the ratio of tangential rotational velocity to translational velocity for dimers and trimers. For each dimer, $r$ is the diameter of one of its constitutive rods; for each trimer, $r$ is $(1 + 2/\sqrt{3})$ times the radius of one its constitutive rods. $\omega$ is the rotational velocity of a particle at a particular time, while $v$ is its translational velocity at the same time. Only particles that have moved at least one monomer radius $R$ from its initial position are considered. For a large majority of dimers and trimers, motion is primarily associated with translations rather than rotations. Rotational motion is not measured for monomers.}
\end{figure}

Over a triangle that is derived from particle positions and Delaunay triangulation, one of the same triangles used in calculating $Q_k$, we calculate $J_2$ using the constant strain triangle formalism~\cite{Cook1974}. We must first note that for all results related to $J_2$ discussed, unless specified, we are focusing on the ``molecular" form of triangulation as defined in Section~\ref{subsec:Structure}. As such, we are treating each point in the triangle as discrete particles, capable of moving independent of each other. For the three particles that make up the triangle, we note the velocity of each particle, each having horizontal component $v_x$ and vertical component $v_y$. Subtracting off the average velocity of the three particles, which is prescribed to the center of mass of the triangle, we determine the local strain tensor $e$, 
\begin{equation}
	\left(\begin{array}{c}
		v_x(x,y)-v_{x,CM} \\
		v_y(x,y)-v_{y,CM}
	\end{array}\right)
	=
	\left(\begin{array}{cc}
		e_{11} & e_{12} \\
		e_{21} & e_{22}
	\end{array}\right)
	\left(\begin{array}{c}
		x \\
		y
	\end{array}\right),
\end{equation}
where $x$ and $y$ are Cartesian coordinates relative to the center of the triangle. One way to conceptualize this formalism is to place pins at the particle centroid locations, with some sort of continuous triangular mesh in the middle. We can deform the mesh by moving the pins relative to each other, causing it to stretch, deform, rotate, or some combination thereof. For this study, we choose not to incorporate particle rotations, which are certainly present, into the formulation of this strain tensor, in part because they substantially complicate the local strain tensor. Also, Fig.~\ref{fig:cdf_rotFrac} indicates that particle motion within dimer and trimer pillars is primarily attributed to translational motion, so a simple strain based on translations alone is likely sufficient to characterize local deformations in this study. Given that acetal rods are slippery compared to the acrylic substrate, grain-grain friction is likely too small to induce rotational velocities that are comparable to translational velocities.  Further studies could explicitly incorporate particle rotations as a component of a more complex local strain, especially in systems of highly frictional grains.

From the empirical local strain tensor $e$, we deduce the symmetric strain tensor $\varepsilon$,
\begin{equation}
	\varepsilon_{ij} = \frac{e_{ij}+e_{ji}}{2}.
\end{equation}
This local linear strain tensor has a number of invariant quantities that characterize the local relative motion of the grains; for instance, the trace defines the dilation rate. We choose to focus on the deviatoric strain rate, $J_2$, as a measurement of the amount of local plastic deformation, 
\begin{equation}
	J_2 = \frac{1}{2}\sqrt{(\varepsilon_{11}-\varepsilon_{22})^2+4\varepsilon_{12}^2}.
\end{equation}

With $J_2$ defined, we should note that there exist other metrics that can fill the role of quantifying local plastic deformation in a similar fashion, e.g., $D^2_{min}$~\cite{FalkLangerPRE1998}. For this study, we choose to focus on $J_2$ for a few reasons. First, as we shall soon discuss, we would like to make direct comparisons with stress, which is a single measurement made at each time point. Thus, we would like to select a kinematic quantity that can also be prescribed to a single time. By definition, $D^2_{min}$ requires the choice of a substantial time interval over which to measure plastic displacements. $J_2$ is calculated from velocities obtained through differentiation over a small time interval as described in Section~\ref{sec:methods}, so it can naturally coincide with the same time point of a stress measurement. Second, $D^2_{min}$ requires the choice of an interaction cutoff length, while $J_2$, derived from Delaunay triangulation, requires no such cutoff. Third, while calculated over the area surrounding a single grain, $D^2_{min}$ is assigned to each individual grain. $J_2$ is rather assigned to a region connected to three grains, so it is a slightly coarse-grained measurement, in line with the approach of established avalanche models~\cite{DahmenNatPhys2011}.

Note that $J_2$ is a strain rate, so it has dimensions of inverse time. $J_2$ is thus scaled relative to the inverse time required to compress the pillar by one large monomer radius, $v_c/R$. This is done for all grain shapes, which have distinct sizes but are all undergoing the same global strain rate. For the sake of comparisons with the global measurement of stress, we take the ensemble average $\langle J_2 \rangle$ as a way to quantify the total amount of plastic deformation throughout the system. We exclude stationary triangles, those that have moved less than a large monomer radius, from the ensemble average $\langle J_2 \rangle$. To further confirm the utility of $J_2$ in quantifying plastic strain, we consider its relationship with stress fluctuations discussed previously in Sections~\ref{subsec:Strength} and~\ref{subsec:Avalanches}.

\begin{figure}
 	\includegraphics[width=0.80\linewidth]{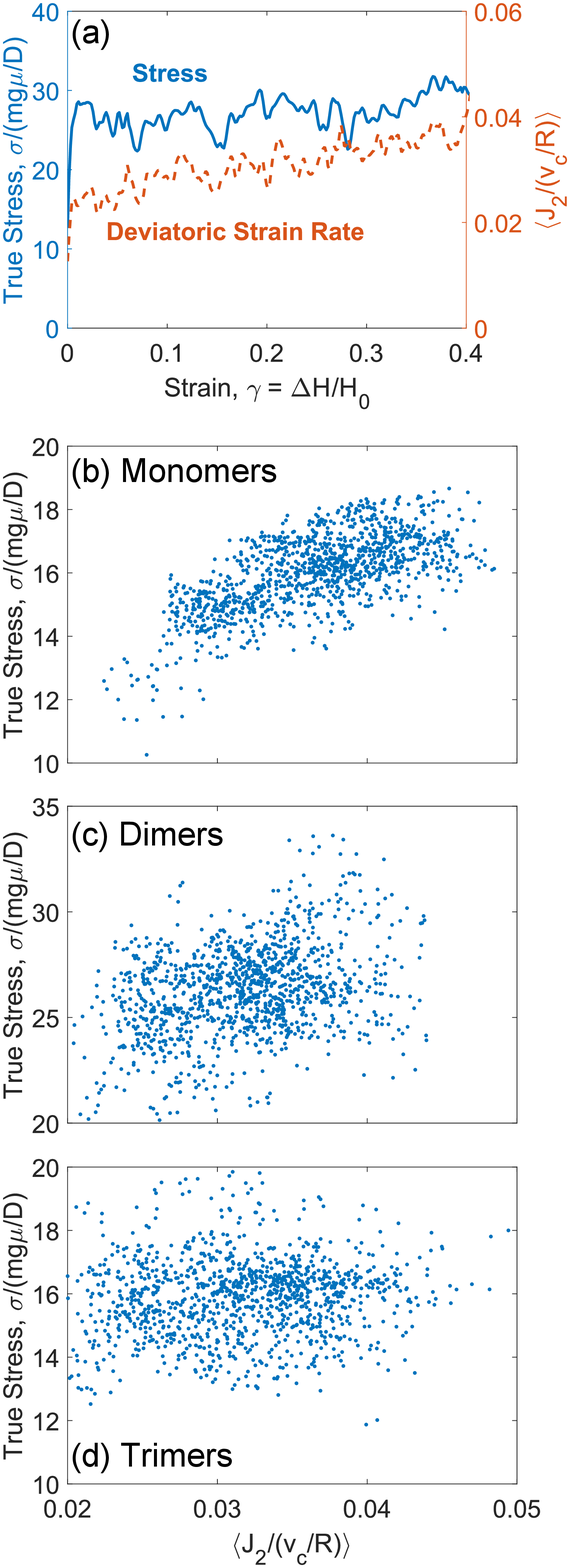}
	\caption{\label{fig:rawStressStrain_stressAvgStrain} (a) For the same trial of dimers as Fig.~\ref{fig:sampleStressStrain}, a plot of both stress $\sigma$ and the average deviatoric strain rate $\langle J_2 \rangle$ throughout the pillar. Note that many of the peaks and troughs of $\langle J_2 \rangle$ correspond to those of $\sigma$.  (b)-(d) $\sigma$ plotted versus $\langle J_2 \rangle$ using data from all recorded trials, demonstrating varying levels of correlation between the global material response and local deformation profile for the different shapes. In (b) and (c), we see a general positive correlation for monomers and dimers, suggesting that local deformations are a primary mechanism for stress relaxation. In (d), we see this trend is mostly absent for trimers, suggesting that the particles are now so large that stick-slip motion due to body friction between particles and the substrate is contributing more to fluctuations in stress.}
\end{figure}

In Fig.~\ref{fig:rawStressStrain_stressAvgStrain}(a), we see that peaks and troughs of $\sigma$ and $\langle J_2 \rangle$ over the course of a single dimers trial generally correlate with each other. We explore the relationship between $\sigma$ and $\langle J_2 \rangle$ further in Fig.~\ref{fig:rawStressStrain_stressAvgStrain}(b)-(d), by plotting the two quantities from all trials directly against each other. We see that in the case of monomers, in Fig.~\ref{fig:rawStressStrain_stressAvgStrain}(b), there is a general positive correlation between the two quantities. This is indicative of particle rearrangements within these pillars as significantly contributing to the presence of avalanches. That is, as the pillar is driven, stress builds up for some period of time. These periods of large $\sigma$ tend to be associated with large $\langle J_2 \rangle$, suggesting that built up levels of stress are subsequently relaxed away by particle rearrangements within the pillar. Fig.~\ref{fig:rawStressStrain_stressAvgStrain}(c) shows that the correlation between $\sigma$ and $J_2$ for dimers is less pronounced. In Fig.~\ref{fig:rawStressStrain_stressAvgStrain}(d), we see that trimers do not exhibit much of a correlation between $\sigma$ and $\langle J_2 \rangle$. The Pearson correlation coefficient $\rho$ of the three sets of data shown in Fig.~\ref{fig:rawStressStrain_stressAvgStrain}(b-d) are as follows: monomers, $\rho = 0.66$, dimers, $\rho = 0.33$, and trimers, $\rho = 0.087$. Since the dimers and trimers are incrementally more massive than the monomers, the shifts in correlation may be due to the fact that stick-slip motion between individual particles and the substrate becomes more prevalent due to body friction. Still, the fact that we see correlations for monomers, and even dimers to a degree, is indicative that avalanches, derived from either global stress or local strains, can be applied to systems of either symmetric or elongated particles. 

\subsection{Structure-Dynamics Connections}
\label{subsec:Connections}

Finally, we discuss connections that can be made between our previous results to quantify both local structure and local dynamics. As $Q_k$ quantifies local under- and over-packing relative to the surrounding neighborhood of a localized region, we can also measure the deviatoric strain rate $J_2$ in a way to highlight regions that are deforming relative to its surroundings~\cite{RieserThesis}. In this way, we emphasize rearrangements that are highly localized as well as rigid areas that are adjacent to shear bands. $J_{2,rel}$, a \emph{relative} deviatoric strain rate, for a given triangle is defined by the difference of its $J_2$ and the average of its neighbors,
\begin{equation}
\label{eq:relJ2}
	J_{2,rel} = J_2 - \langle J_{2,neighbors} \rangle.
\end{equation}
Neighboring triangles are defined to be those which share at least one vertex, i.e., particle, with a given triangle. 

\begin{figure*}
 	\includegraphics[width=\textwidth]{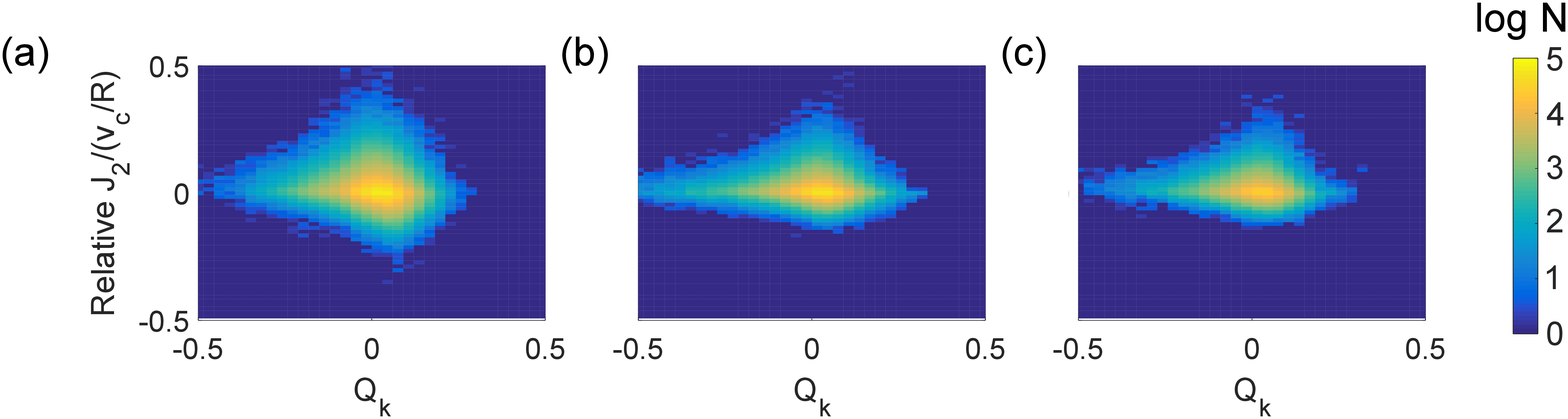}
	\caption{\label{fig:Strain_parts_Raw} Bivariate histograms of $J_{2,rel}$, defined in Equation~\ref{eq:relJ2}, versus $Q_k$ for (a) monomers, (b) dimers, and (c) trimers. The plots are colored by the log (base 10) of counts within each bin.}
\end{figure*}

In Fig.~\ref{fig:Strain_parts_Raw}, we show the raw bivariate histograms for values of $Q_k$ and $J_{2,rel}$, highlighting the amount of spread in $J_{2,rel}$ at each value of $Q_k$. In general, one should expect under-packed regions are more likely to undergo strain than over-packed regions, since a void region with open space can collapse more easily. Meanwhile, over-packed regions are more constrained by its neighbors, so those can be expected to be less likely to strain. However, one must note that individual structural metrics can be poor predictors of particle rearrangements~\cite{CubukSchoenholzPRL2015}. Indeed, it is difficult to observe any trend in Fig.~\ref{fig:Strain_parts_Raw}, although a slight negative correlation between $J_{2,rel}$ and $Q_k$ may be visible. 

\begin{figure}
 	\includegraphics[width=\linewidth]{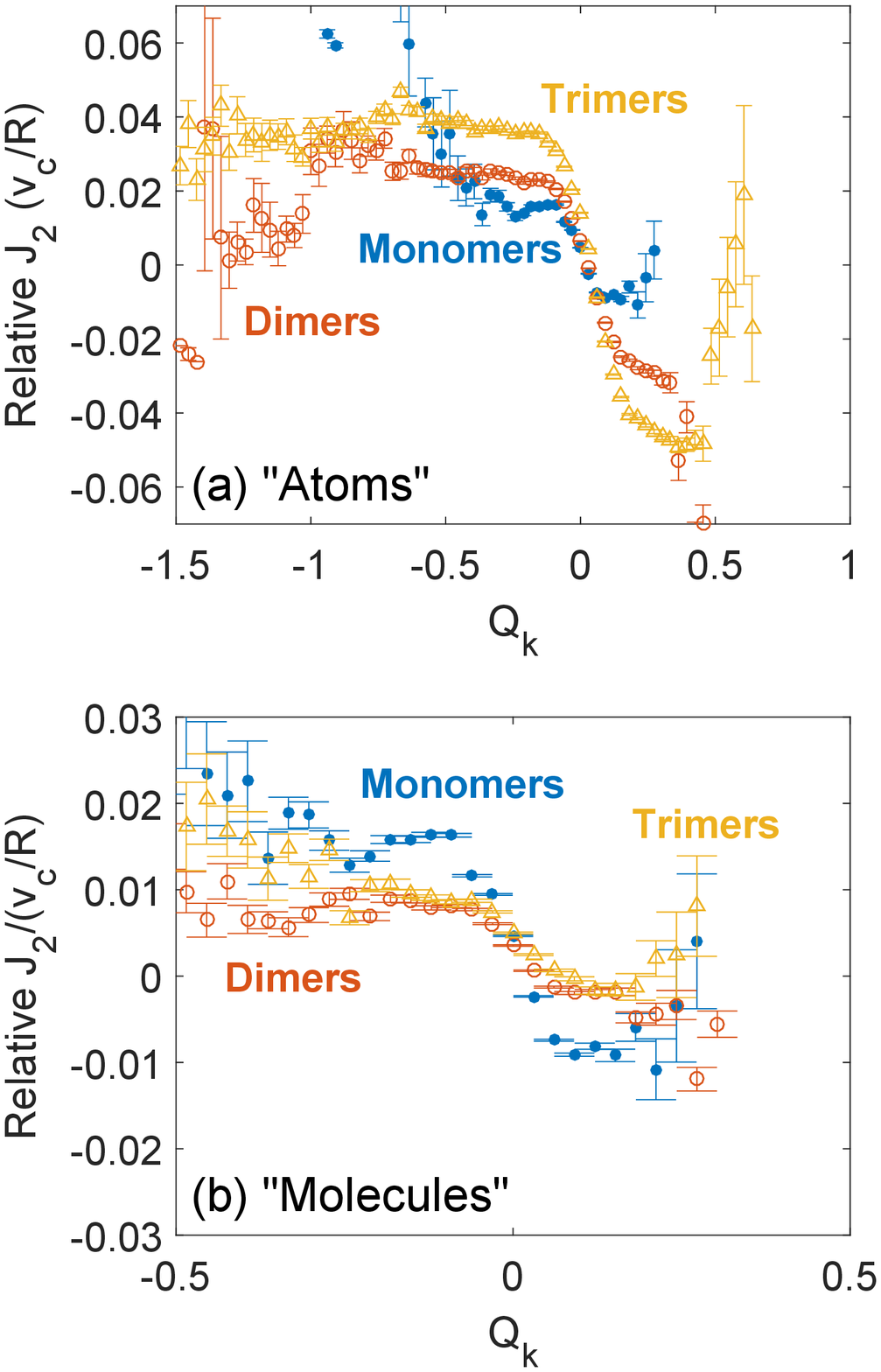}
	\caption{\label{fig:Strain_atomsMolecs} Binned averages of $J_{2,rel}$ versus $Q_k$ for all particle shapes as measured in (a) ``atoms'' and (b) ``molecules.''  The error bars are standard errors of the mean of $J_{2,rel}$ values within each $Q_k$ bin. In general, there is a negative correlation between the two, aside from dilatancy effects at maximally packed regions (max $Q_k > 0$) for monomers and trimers. However, dimers do not exhibit this dilatancy effect, while regions of dimers that are highly under-packed ($Q_k \lesssim -1.0$ in (a), $Q_k \lesssim -0.3$ in (b)) show a lower likelihood to deform compared to monomers and trimers.}
\end{figure}

To specify an overall trend of $J_{2,rel}$ versus $Q_k$, we bin the data in Fig.~\ref{fig:Strain_parts_Raw} by intervals in $Q_k$ and take averages of corresponding values of $J_{2,rel}$ to generate Fig.~\ref{fig:Strain_atomsMolecs}(b). Here, the trend of $J_{2,rel}$ with $Q_k$ is much more apparent. The vertical error bars represent standard deviation of the mean $J_{2,rel}$ within each $Q_k$ bin. Given that bins near $Q_k = 0$ contain $\mathcal O(10^5)$ data points, these error bars are vastly suppressed compared to the actual spread in raw data. 

Indeed, we observe a negative correlation between these two quantities, which is approximately linear in the region around $Q_k = 0$. This is in line with the expected trend of how likely under- and over-packed region are expected to deform. That is, voids are readily collapsing, while constrained regions are persisting. The linear dependence is more pronounced among monomers, likely due to the absence of geometrical constraints such as elongation and bumpy surfaces. However, we note two important deviations from this general behavior that occur at the opposite extremes in $Q_k$. 

At highly positive $Q_k$, we see a dramatic upturn in $J_{2,rel}$ for monomers and trimers, indicative of the onset of Reynolds dilatancy~\cite{ReynoldsDilatancy}. The acetal rods are quite hard, so there is a finite limit to how closely they can be packed until they must deform locally. However, dimers do not demonstrate such a dramatic increase, capturing the ability of many dimer pairs to interlock and actually form rigid structures, as demonstrated in Fig.~\ref{fig:Ordered_Dimers}(b). The second notable deviation we see is at highly negative $Q_k$ values, which indicate large voids in the packing, we see a dip in $J_{2,rel}$ toward zero for dimers, indicating that the void regions that form in dimers can actually persist for some time, unlike large voids in monomers and trimers that readily collapse. The dip in $J_{2,rel}$ for discrete dimers in present in Fig.~\ref{fig:Strain_atomsMolecs}(b), but is especially apparent when measured using the ``atomistic'' approach shown in Fig.~\ref{fig:Strain_atomsMolecs}(a). This deviation in the dynamical behavior of dimer ``molecules'' in Fig.~\ref{fig:Strain_atomsMolecs}(b) coincides with a kink in the dimer $Q_k$ distribution shown in Fig.~\ref{fig:Qk_ALL}(b) over the same highly negative $Q_k$ region, starting near $Q_k \sim -0.3$. 


To interpret these results, we note that highly packed regions and large voids are actually quite inter-related. When a large void opens up in the packing, for any particle shape, it is always surrounded by a ring of tightly packed grains. As such, the persistence of a void region requires similar persistence of neighboring over-packed regions. 

\section{Discussion}
\label{sec:Discussion}

In this article, we presented an experimental study into how a granular pillar, acting as a model disordered solid, deforms under uniaxial compression with varying particle shapes created from bonded groups of circular rods. We see that dimers constitute the strongest pillars, the additional strength originating from dimers that align and interlock perpendicular to the compression direction. The capability of horizontally oriented dimers to bear substantial loads can be seen in pillars in which the initial configuration contains dimers that are preferentially ordered horizontally. While dimers are clearly capable of interlocking as a form of inter-particle friction, it would be interesting to investigate elongated convex shapes, such as ellipses, to determine whether pillar strengthening, in addition to other results presented, can be reproduced regardless of convexity. For example, simulated systems with a similar geometry have seen increased material strength that results from increased contact area yielding additional sliding contacts~\cite{AzemaPRE2010}, so it would be informative to perform  experimental tests. Furthermore, when convex shapes are used, perhaps rotational frustration will play a larger role in determining the response of the pillar.

As is the case in other driven amorphous systems, the stress response of our granular pillars exhibits stress relaxations, or avalanches, over time. Furthermore, we see that particle shape does not affect the exponent of the power law distribution that tends to be representative of avalanche sizes within a wide range of amorphous systems. We do see that the more frictional shapes, dimers and trimers, allow for larger avalanches. While the local mechanisms for avalanches are seemingly unaffected, particle shapes do affect the local threshold stress that precedes local deformation, possibly by way of increased interparticle friction for our concave shapes.

We also characterize local structure within the pillar using the previously defined structure metric $Q_k$~\cite{RieserPRL2016} to highlight local packing anisotropies, which manifest as both voids and compacted regions. When $Q_k$ is computed based on the positions of composite dimers or trimers, rather than component circular rods, we see that the $Q_k$ metric retains its Gaussian-like characteristic. This indicates promise in the utility of $Q_k$ as a randomly distributed measure of local free area in disordered packings of arbitrary particle shape and size. As previously stated, convex shapes would also prove to be a valuable test for $Q_k$, since voronoi tessellation would require curved facets and a more complex computation of local free area~\cite{SchallerEPL2015}.

Finally, we measure local strain rates within the pillar and draw correlations with local structural anisotropies. For all shapes we note a general average trend that under-packed regions tend to be more likely to rearrange or undergo strain. Meanwhile, an over-packed region is less likely, on average, to deform, up to the limit where maximally packed grains must undergo dilatancy~\cite{ReynoldsDilatancy}. While this trend is generally true for all shapes, there is a clear deviation for highly under-packed dimers, which are not as likely to deform. This feature captures the observation that dimers can readily form voids that remain even for large strains. We expect that the aggregation of local rigidity can also play a key role in the global strengthening of the pillar.

Using some of the techniques described in this article, in junction with the machine learning approach introduced in Ref.~\cite{CubukSchoenholzPRL2015}, we are interested in pursuing studies that further connect local structural defects with particle-scale rearrangements of asymmetric particles. A probabilistic description of the likelihood of a region within a material to fail, in junction with structure functions that account for grain shape, may then illuminate strategies to prevent vulnerable local structures from forming in a wide range of disordered solids. While this study focuses mostly on particle trajectories and local structural features, further studies that include force measurements between grains would elucidate local stress capacities and add another consideration that influences whether a region is likely to deform locally. Furthermore, measuring forces between grains would allow for direct experimental comparisons to established models of ductile failure, with uniquely direct knowledge of the material microstructure. As mentioned in Section~\ref{sec:Intro}, ductile failure occurs due to local plastic flow and/or the growth and coalescence of voids~\cite{PINEAU2016424}, both of which are present in the deformation of our pillars. Ductile failure can be modeled using the constitutive GTN model~\cite{Gurson1977,Tvergaard1981,TvergaardNeedleman1984} and modifications thereof. However, these models are governed by a balance between locally applied stresses and yield stresses, both of which are currently inaccessible in the present study. Still, these prospective studies show promise for a more thorough understanding of material failure in disordered solids, and can illuminate methods for mitigating or avoiding catastrophic failure events. 

\section*{Acknowledgments}
We thank J. M. Rieser for technical assistance. This work was supported by NSF grants MRSEC/DMR-1120901 and MRSEC/DMR-1720530.

\bibliography{Dimers_bib}

\end{document}